\documentclass[twocolumn,english,pra,showpacs,superscriptaddress,longbibliography]{revtex4-1}
\usepackage{amsfonts}
\usepackage{amsmath}
\usepackage{amssymb}
\usepackage[colorlinks, citecolor=blue,linkcolor=red]{hyperref}
\usepackage{color}
\usepackage{hyperref}
\usepackage{textcomp}
\usepackage[rightcaption]{sidecap}
\usepackage{subfigure}
\usepackage[rightcaption]{sidecap}
\usepackage{graphicx}
\usepackage{lipsum}

\begin{document}
\title{$\mathcal{PT}$-Symmetry induced Bi-Stability in Non-Hermitian Cavity Magnomechanics}
\author{Chaoyi Lai}
\affiliation{Department of Physics, Zhejiang Normal University, Jinhua 321004, China.}
\author{Shah Fahad}
\affiliation{Department of Physics, Zhejiang Normal University, Jinhua 321004, China.}
\author{Kashif Ammar Yasir}
\email{kayasir@zjnu.edu.cn}\affiliation{Department of Physics, Zhejiang Normal University, Jinhua 321004, China.}
	%%%%%%%%%%%%%%%%%%%%%%%%%%%%%%%%%%%%%%%%%%%%%%%%%%%%%%%%%%%%%%%%%%%%%%%%%%%%%%%%%%%%%%%%%%%%%%%%%%%%%%%%%%%%%%%%%%%%%%%%%%%
\setlength{\parskip}{0pt}
\setlength{\belowcaptionskip}{-10pt}
\begin{abstract}
We study the steady-state non-Hermitian magnomechanical system driven by a transverse magnetic field directly interacting with YIG sphere and excites cavity magnons and photons. To make the system non-Hermitian, we use a traveling field directly interacting with magnons generating gain to the system. We start by illustrating PT-configuration of the system, which contains two PT broken region around exceptional point and PT protected region along the axis of exceptional point. Late, we discover that the numbers of cavity photons and magnons show bistable behavior depending upon the PT configuration, which becomes more significant as the values of the magnon-photon coupling and traveling field strength increases. We illustrate that steady-state photon only shows bistable behavior when the system in in lossy PT broken configuration, means strength of traveling field is less than the magnon-photon coupling. Otherwise, it will just contain a single stable state because of bistability suppression with gain in the system, which is unlike with any other investigation in this direction. Further, a larger magnon-photon coupling increases photon intensity and decreases magnon intensity, because of photon and magnon energy exchange, leading to enhanced photon bistablity and decreased magnon bistability. However, in case of increasing strength of traveling field, both photon as well as magnon bistability is appeared to be decreasing. We also study the steady-state effective potential of the system and illustrate the occurrence of bistability with nonlinear interactions between contour trajectories, which similarly depends on the PT broken configuration of the system.
\end{abstract}
%\pacs{42.50.Pq, 42.50.Gy, 67.85.Hj, 71.70.Ej}
\date{\today}
\maketitle
	
\section{Introduction}
Over the years, mechanical oscillators with radiation pressure have gained great attention as transducers mediating coherent signals across various systems \cite{Aspelmeyer_2014_Cavity}.
In this stunning direction, a variety of methods are discovered and widely used to facilitate the weak to strong coupling regimes between phonons and optical or microwave photons, such as piezoelectric force \cite{Bochmann_2013,Fan_2015_Cascaded}, radiation pressure mediated optomechanical cavities \cite{Kippenberg_2007,Mo_2008_Harnessing,Park_2009_Resolved,Weis_2010,Safavi-Naeini_2011,Hill_2012,Carmon_2005_Temporal,Kippenberg_2005_Analysis,Rokhsari_2005,Arcizet_2006,Gigan_2006,Schliesser_2006_Radiation,Yasir_2016,Yasir_2022} 
and electrostatic forces induced motions \cite{Teufel_2011,Andrews_2014,Bagci_2014}. Such kind of understanding of hybrid system interactions contribute to the rapid evolution of diverse electro-cavity and optomechanical systems \cite{Aspelmeyer_2014_Cavity}.
In ferromagnetic materials, magnon is the collective excitation of the magnetization, which can be controlled by the applied magnetic field, where the magnons and phonons are coupled through the magnetostrictive force, which provides a new way to carry different information \cite{Serga_2010,Lenk_2011,Chumak_2015}. 

Meanwhile, ferromagnetic system, like yttrium-iron-garnet (YIG), provides opportunity for strong light-matter interaction resulting in the kittle like modes that generate strong coupling with cavity photons \cite{Kittel_1948,Huebl_2013_High,Tabuchi_2014_Hybridizing,Zhang_2014_Strongly,Goryachev_2014_High,Bai_2015_Spin}.
The magnetic dipoles interaction induced by magnetostrictive forces in presence of cavity mode give birth to highly interactive magnetic entities known as magnons \cite{Tabuchi_2014_Hybridizing,Hubel_2013_High}.
YIG sphere serves as an effective magnomechanical resonator, which is integrated with cavity quantum electrodynamics to form a hybrid magnomechanical (CMM) systems.
The CMM system has been proven to be best tool to test various semi-classical and quantum phenomena, for example; magnon bistability \cite{Wang_2016_Magnon,Hyde_2018_Direct}, cavity-magnon polaritons \cite{Cao_2015_Exchange,Yao_2015_Theory}, magnon dark mode \cite{Zhang_2015_Magnon} and non-Hermitian exotic properties \cite{Harder_2017,Wang_2018}. 
Further, the two-drive approach with experimental feasibility of the generic CMM system provides a practical route for the development of tunable phonon lasering devices with properties, such as, low-threshold, high-gain, and narrow-linewidth, which are based on the platform of CMM system \cite{Zhang_2023_Generation}.
S.N. Huai et al, \cite{Huai_2019_Enhanced} have studied that the CMM system is a platform for simultaneously exploring PT-symmetric-like paradigms in the microwave field, and it is a good candidature to control the microwaves on both first- and high-order side-bands. There are many other interesting phenomena have been studied in CMM system, such as the entanglement of magnon-photon-phonon \cite{Li_2018_Magnon}, dynamical multistability of cavity magnomechanics \cite{Wenlin_2023_Nonlinear}, steering \cite{Tan_2021_entanglement}, coherence \cite{Qiu_2022_Controlling}, and squeezing \cite{Li_2019_Squeezed}. X. Zhang et al. \cite{Zhang_2016_Cavity} investigated an intriguing CMM system and studied the magnomechanically parametric amplification, magnomechancally induced transparency/absorption and triple resonance based on the coherent coupling between photons, magnons and phonons. Although there are many interesting investigations have been done in this direction, but the non-Hermitian aspects of magnonic cavities are still need to be explored.

In non-Hermitian systems, the parity-time (PT) symmetry plays a key role, which make non-Hermitian systems crucial in order to study unconventional phenomenon of physics and is the subject of increasing investigations. Recently, W.B.Rui et.al \cite{Rui_2019_semimetals} studied the PT-symmetric Dirac semimetals on periodic lattices perturbed by general non-Hermitian potentials.
S.K Gupta \cite{Gupta_2020} discussed the developments in the field of complex non-Hermitian physics based on PT-symmetry in various physical settings along with including key concepts and background.
S.Xia et.al. \cite{Xia_2021} theoretically examined the interactions between exceptional points (EP) and non-Hermitian topological modes. This provides a new platform to study the nonlinear effects in non-Hermitian topological systems and discussed the single-channel control of global PT-symmetry through local nonlinearity.
B. Wang. et.al, \cite{Wang_2022_laser} demonstrated that the stimulated emitted number of magnons coherently increased above a threshold driving power in the PT-symmetric regime, when the magnon frequency resonant with the splitting of optical superdomes, indicating the occurrence of the magnon laser. 

In this paper, we theoretically investigate the steady-state behavior of hybrid non-Hermitian magnomechanical system containing a YIG sphere directly driven by coherent magnetic field source exciting magnonic modes. The YIG sphere also scatters photons to the cavity depending upon the external driving field, which later excite mechanical phonons. We use another traveling field directly interacting with the YIG to make system non-Hermitian. First, we illustrate the PT-symmetric behavior the non-Hermitian system, and we show that the both photon and magnon numbers contain steady-state bistable behavior in relation to the PT-symmetric configuration. From calculations, we illustrate that the both magnon-photon coupling and non-Hermitian parameter alter the bistable dynamics of photon number as well as magnon number, providing opportunity for tunability. We also discuss the dependence of bistable behavior on PT broken and unbroken configurations with traveling field strength. Further, we numerically investigate the steady-state effective potential of the system and find that it contains similar nonlinear effects like bistability, which again crucially depends on non-Hermitian strength as well as on the magnon-phonon coupling. 
\begin{figure}[tp]
	\includegraphics[width=8cm]{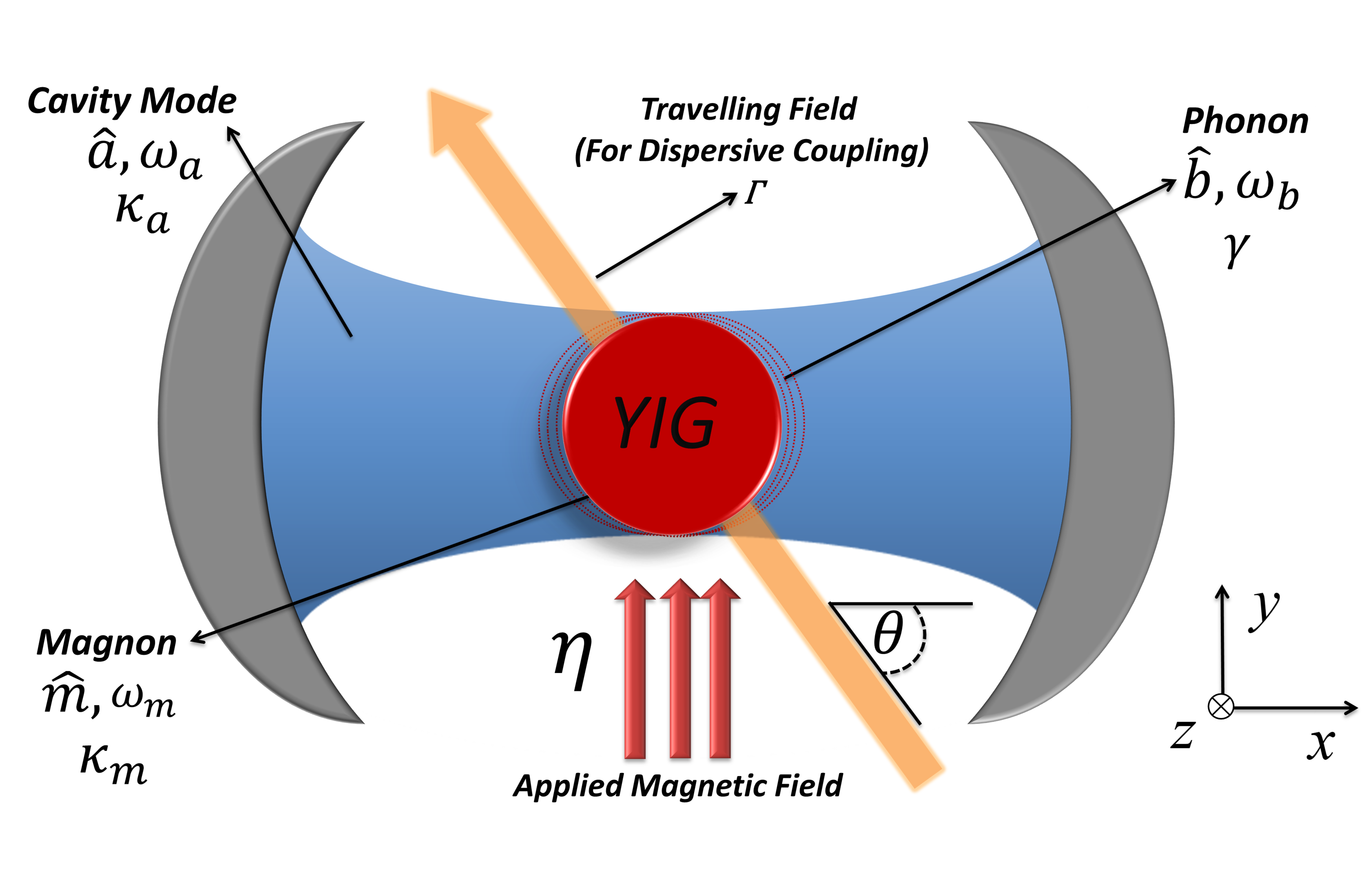}
	\caption{The schematic description of a dissipative magnon-optical cavity system, in which the cavity is driven by a  Magnetic field $\eta$ along $y$ -axis.}
	\label{fig1}
\end{figure}\label{Fig1}

The manuscript is organized as follows: Section \ref{sec2} describes the system mechanism and the mathematics. Section \ref{sec3} demonstrates the broken and unbroken dynamics of the PT-symmetry of the system. While the section \ref{sec4} illustrates the occurrence of magnonic and photonic multistability in the system. Finally, section \ref{sec4} presents the conclusion of the work. 

\section{Non-Hermitian Magnomechanics: A System Description}\label{sec2}
The system consists of a high-$Q$ Fabry–Pérot cavity of length $L\approx12.5\times10^{-3}\mathrm{m}$ containing a ferromagnetic YIG sphere, usually in the range of $10^2 \mathrm{\mu m}$ to $1\mathrm{mm}$ in diameter, as illustrated in  Fig.\ref{fig1}. The system is driven by a $10\mathrm{G}$ bias orthogonality (perpendicular to the cavity axis) $B_0$ magnetic field directly interacts with YIG with strength $\vert\eta\vert=\sqrt{P\times\kappa/\hbar\omega_{B}}$, where $P=0.0164\mathrm{mW}$ represents the power of source and $\omega_{B}=3.8\times2\pi\times10^{14}\mathrm{Hz}$ is the frequency of the applied magnetic field. The interaction of magnetic field excites magnons $\hat{m}$ on the surface of YIG sphere, which then results in coupling between magnons and photonic $\hat{a}$ cavity mode excited by the photon scattering from YIG. These scattered photons build a strong cavity mode, because of the high-$Q$ factor of the cavity mirrors, oscillating at $\omega_{c}\approx1.9\times2\pi \mathrm{GHz}$ with $\Delta_c=\omega_B-\omega_c$. The cavity mode later generates phonon $\hat{b}$ modes around YIG because of optomechanical interactions of cavity photons. Because of the high-$Q$ factor the collective cavity leakage is considered as $\kappa\approx 1\times2\pi\mathrm{MHz}$, which should satisfy the condition $Ng_0^2/\Delta_a>>\kappa$ for strong coupling regime. To make our findings experimentally possible, we selected mentioned parameters for calculations from currently available state-of-the-art experimental setups \cite{Kittel_1948,Huebl_2013_High,Tabuchi_2014_Hybridizing,Zhang_2014_Strongly,Goryachev_2014_High,Bai_2015_Spin}.

The non-Hermitian Hamiltonian offer a different insight into the dynamics of classical and quantum systems connected non-conservatively to a surrounding environment. The Hamiltonian for the proposed system can be divided into three parts,
\begin{equation}
\hat{\mathcal{H}}=\hat{\mathcal{H}}_{0}+\hat{\mathcal{H}}_{int}+\hat{\mathcal{H}}_{dr},\label{General equation}
\end{equation}
where, $\hat{\mathcal{H}}_{0}$ is the hermitian part of the Hamiltonian containing solo energies of associated subsystems. $\hat{\mathcal{H}}_{int}$ contains the interaction between these associated subsystems and, finally, $\hat{\mathcal{H}}_{dr}$ contains the non-Hermitian part of the system, which consists of the energies interaction associated with traveling field. These parts of the Hamiltonian can be expressed as,
\begin{eqnarray}
\hat{\mathcal{H}}_{0}&=& \omega _{c}\hat{a}^{\dagger}\hat{a}+\omega _{m}\hat{m}^{\dagger}\hat{m}+\omega_{b}\hat{b}^{\dagger}\hat{b} ,\\
\hat{\mathcal{H} }_{int}&=&G_{a}(\hat{a}+\hat{a}^{\dagger })(\hat{m}+\hat{m}^{\dagger })+G_{b}\hat{m}^{\dagger }\hat{m}(\hat{b}^{\dagger}+\hat{b})\nonumber\\
&&-\mathrm{i}\eta(\hat{m}^{\dagger}\mathrm{e}^{-\mathrm{i}\omega_{0}t } -\hat{m}\mathrm{e}^{\mathrm{i}\omega_{0}t }) ,\\
\hat{\mathcal{H}}_{dr}&=&-\mathrm{i}\Gamma \mathrm{e}^{\mathrm{i}(\delta t+\theta) }(\hat{a}+\hat{a}^{\dagger })(\hat{m}+\hat{m}^{\dagger }).\label{APT-H}
\end{eqnarray}
Here $\omega_{c}$, $\omega_{m}$, and  $\omega_{b}$ are the frequencies of the photonic, magnonic, and phononic modes, respectively, under assumption $\hbar=1$.
$\hat{a}(\hat{a}^{\dagger})$ represents the annihilation (creation) operators for photons, while $\hat{m}(\hat{m}^{\dagger})$ represents those for magnons. 
$G_{a}$ \cite{Lachance-Quirion_2019,Li-Yi_2020} is the coupling between photons and magnons, and can be defined as $G_a=\sqrt{2}(\omega_{c}/L)x_{m}$, where $x_{m}=\sqrt{\hbar/2m_m\omega_{m}}$ is the zero point motion of magnon having stationary mass $m_m$ and frequency $\omega_{m}$. Similarly, the coupling between magnons and corresponding phonon can be defined as $G_b=\sqrt{2}(\omega_{c}/L)x_{b}$ with $x_{b}=\sqrt{\hbar/2m_b\omega_{b}}$ corresponds to the zero point motion of optomechanical phonons with mass $m_b$ and frequency $\omega_{b}$ \cite{Potts_2021_Dynamical}. Further, $\eta$ is the strength of applied magnetic field as discussed earlier. The last part of the Hamiltonian contains the interactions of externally applied traveling field which directly interacts with cavity magnons. Here $\Gamma$ represents the coupling strength of traveling field with the magnonic mode defining as $\Gamma=\mathcal{\alpha}\sqrt{(\hbar/\omega_m m_m)}$, where $\mathcal{\alpha}$ corresponds to amplitude of the traveling field \cite{Yang_2017_Anti,Zhao_2020_Observation,Elyasi_2020_Resources,Barman_2021_Roadmap,Hurst_2022_Non,Li_2020_Hybrid}. $\delta$ and $\theta$ represent the frequency and angle between cavity axis and traveling field. 
\begin{figure}[tp]
	\includegraphics[width=9cm]{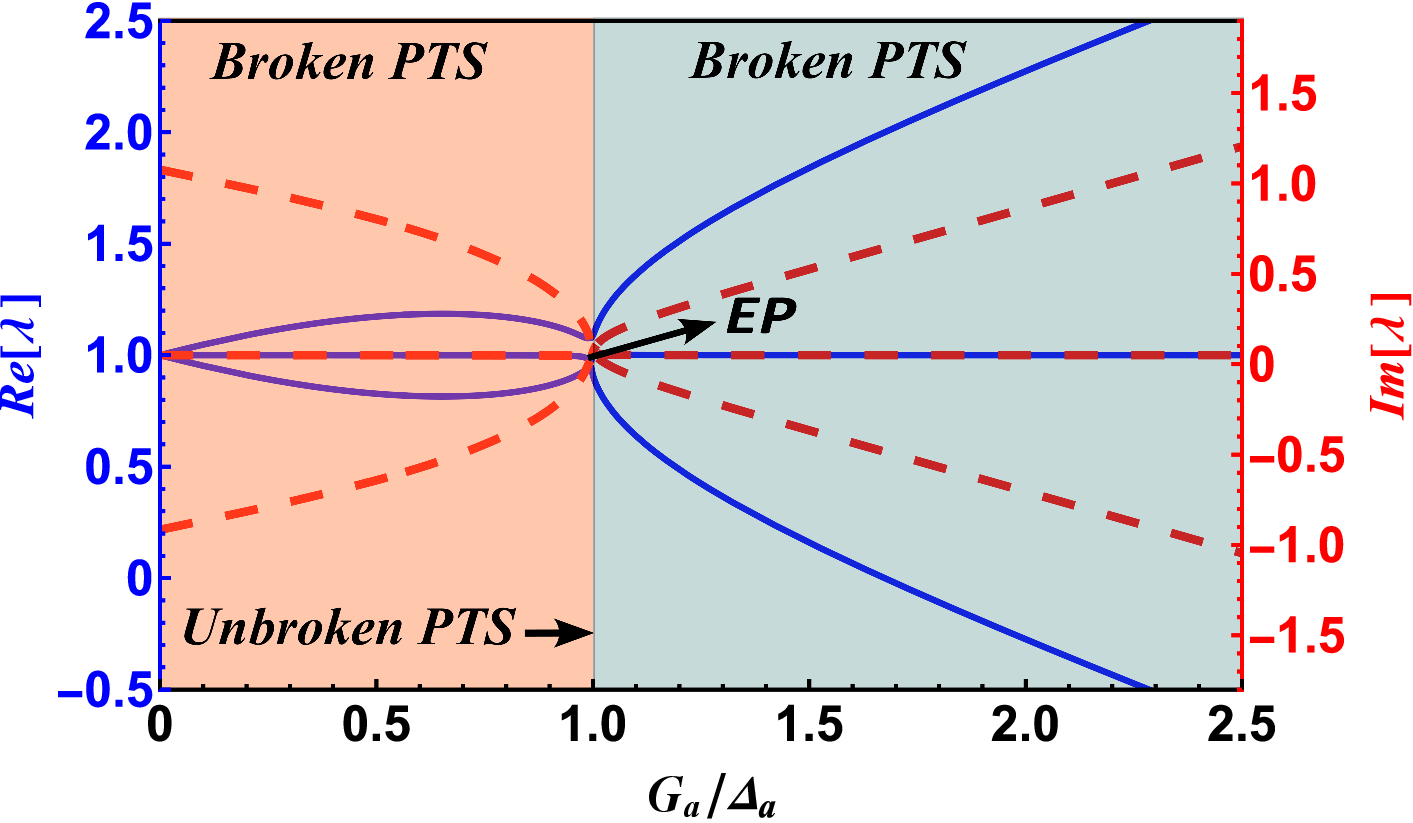}
	\caption{Eigenenergy spectrum versus the normalized magnon-photon coupling rate $G_{a}/\Delta_a$. Here blue curve corresponds to the real part $Re[\lambda]$ and red dotted curve represents imaginary part $Im[\lambda]$ of the eigenvalues, when incident angle of traveling field is $ \theta=\pi $, and non-Hermitian parameter $\Gamma/\Delta_m=1$. Here it should be noted that the reason behind choosing the particle angle is that the system is only PT-symmetric and stable at this angle, see ref\cite{Yasir_2024} The other parameters we choose for the numerical calculation are $ \kappa_m/\Delta_m=1, G_{mb}/\Delta_b=2, \gamma_b/\kappa_m=.01, \Delta_m/\kappa_m=1$ and $\Delta_b/\omega_b=1 $} 
	\label{fig2}
\end{figure}

After this, we apply the rotational approximation \cite{Yasir_2017_Spin,Li_2018_Magnon,Wenlin_2016_Parity}, which will allow us to redefine $(\hat{a}+\hat{m}^{\dagger })(a^{\dagger }+\hat{m})\approx(\hat{a}\hat{m}^{\dagger }+\hat{a}^{\dagger }\hat{m})$, and consider the effects of traveling field in form of neglecting time and frequency component of traveling field $\mathrm{e}^{\mathrm{i}(\delta t+\theta) }\approx\mathrm{e}^{\mathrm{i}\theta }$. The total Hamiltonian will now read as,
\begin{equation}\begin{split}
\hat{\mathcal{H}}&= \Delta _{c}\hat{a}^{\dagger}\hat{a}+ \Delta _{m}\hat{m}^{\dagger}\hat{m}+ \omega_{b}\hat{b}^{\dagger }\hat{b}\\
&+(G_{a}-\mathrm{i}\Gamma \mathrm{e}^{\mathrm{i}\theta  } )(\hat{a}\hat{m}^{\dagger }+\hat{a}^{\dagger }\hat{m}) +G_{b}\hat{m}^{\dagger }\hat{m}(\hat{b}^{\dagger}+\hat{b}) \\
&+\mathrm{i}\eta(\hat{m}^{\dagger} -\hat{m} ) .
\end{split} \label{H_tot}
\end{equation}
From Hamiltonian, one can note that it consists of three subsystems, photon, magnon and phonon. As, the phonons are least coupled to the system and we are more interested in the behavior of magnons and photon, so, in following study will consider the solo dynamics of phonons and will keep ourselves limited to the behavior of photons and magnons. However, the coupling phonon with system plays an important rules, especially in governing behavior of PT symmetry and revealing the presence of a third-order exceptional point (EP) \cite{Jing_2017, Zhitao_2024}, which we will discuss later in this article.

In order to govern the time-dynamics, we use Langevin equations approach and calculated the equation of motion corresponding to each associated subsystem.
\begin{eqnarray}
\dot{\hat{a}}&=& -\mathrm{i}[ ( \Delta_{c}-\mathrm{i}\kappa_{a})\hat{a}+(G_{a}-\mathrm{i}\Gamma \mathrm{e}^{\mathrm{i}\theta } )\hat{m}] ,\\
\dot{\hat{m}}&=& -\mathrm{i}[ (\Delta_{m}-\mathrm{i}\kappa_{m} )\hat{m}  +(G_{a}-\mathrm{i}\Gamma \mathrm{e}^{\mathrm{i}\theta } )\hat{a}+\mathrm{i}\eta] ,\\
\dot{\hat{b}}&=&-\mathrm{i}[ (\omega_{b}-\mathrm{i}\kappa_{b})\hat{b}+G_{b}\hat{m}^{\dagger}\hat{m} ],	
\end{eqnarray}
$\kappa_{m}, \kappa_{a}$ and $\kappa_{b}$ are the mechanical dissipation of magnon, cavity photon, and phonon, respectively, introduced by the Langevin equations approach. 
\begin{figure*}[tp]
	\includegraphics[width=14cm]{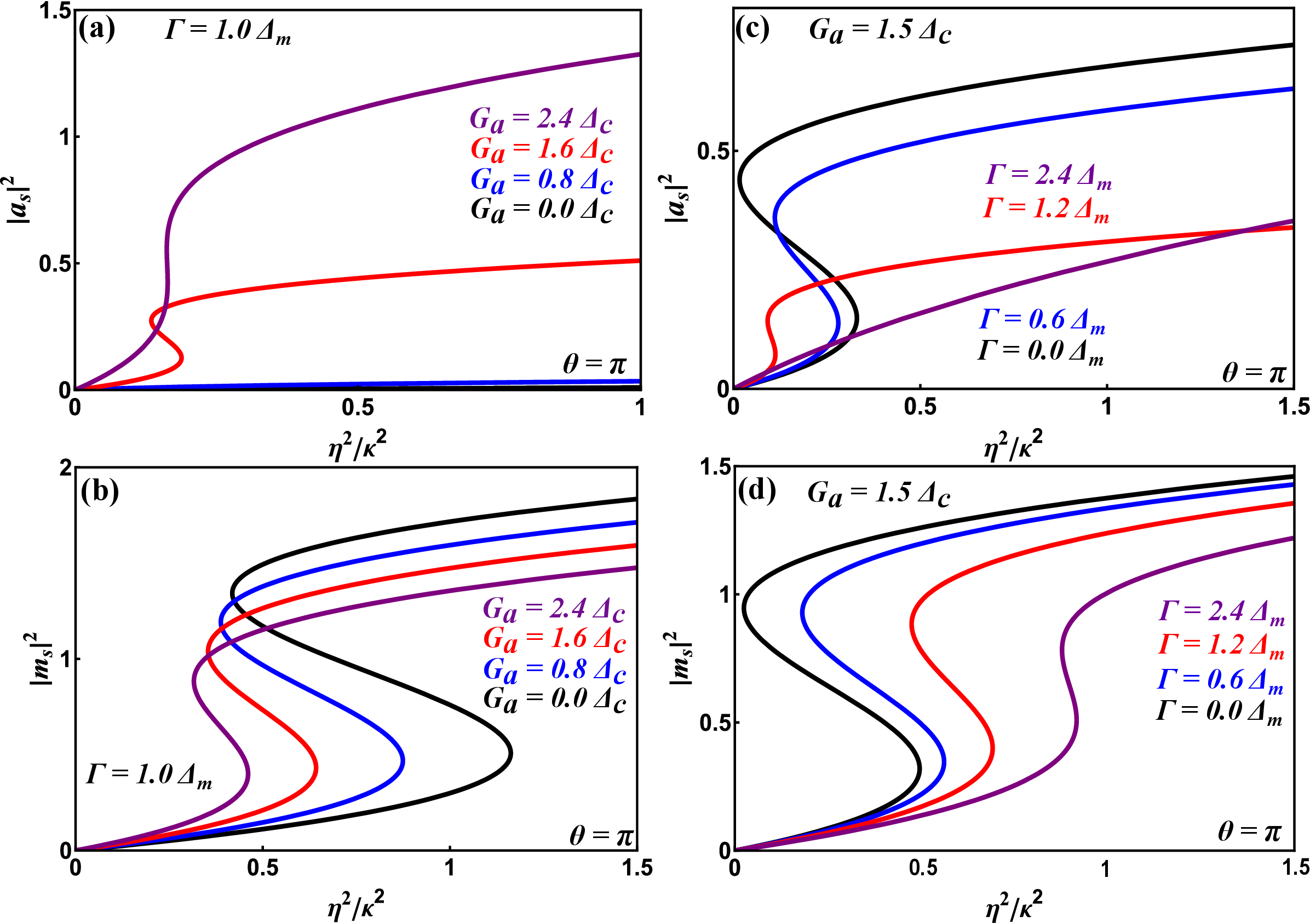}
	\caption{(a) and (b) show the behavior of the steady-state cavity photon number $\left | a_{s} \right |^2$ and  magnon number $\left | m_{s} \right |^2$ as a function of normalized driven magnetic field $\eta^2/\kappa^2$, in which the coupling parameter $G_{a}$, varies from $G_{a} = 0.0\Delta_{m}$,$G_{a} = 0.8\Delta_{m}$,$G_{a} = 1.6\Delta_{m}$, to $G_{a} = 2.4\Delta_{m}$ at constant non-Hermitian parameter $\Gamma_a = 1.0\Delta_m$. (c) and (d) show the steady-state cavity photon numbers $\left | a_{s} \right |^2$  and magnon numbers $\left | m_{s} \right |^2$ with varying $\Gamma$ from $\Gamma = 0.0\Delta_{m}$, $\Gamma = 0.6\Delta_{m}$, $\Gamma = 1.2\Delta_{m}$, to $\Gamma = 2.4\Delta_{m}$ at constant coupling parameter $G_a = 1.5\Delta_c$. Other parameter used in numerical calculation are same as in Fig.\ref{fig2}.}
	\label{fig3}
\end{figure*}

\section{Parity-Time symmetry and Exceptional Points}\label{sec3}
From above mentioned equation of motions, one can drive the effective Hamiltonian for the system (for detailed calculation please see Appendix A), reading as,
\begin{equation}
	\hat{H}_{eff}=\begin{pmatrix}
		\Delta_a+i\kappa_a & G_{a}+i\Gamma e^{i\theta} & 0 \\
		G_{a}+i\Gamma e^{i\theta} & \Delta_m+i\kappa_m & G_{b} \\
		0 & G_{b} & \Delta_b+i\kappa_b  \\
	\end{pmatrix}.
\end{equation}

The effective Hamiltonian clearly possesses three eigenvalues, as illustrated in Fig.\ref{fig2}. It contains the region of Broken PT, exceptional point (EP), and Unbroken PT as a function of normalized magnon-photon coupling. 
For this matrix Hamiltonian, three eigenvalues can be predicted. It should be noted that the PT symmetric behavior of the system has been thoroughly investigated in reference \cite{Yasir_2024} and in this article, we will be focusing on steady-state dynamics of the system. One can easily interpret the stunning relation between the eigenfrequency (eigenvalues or complex frequency) of the magnomechanical system and the dimensionless magnon-microwave coupling rate $ G_{a}/\Delta_a=1 $, where $ \Delta_a $ fixed. When $ G_{a}/\Delta_a=1 $, the eigen-spectrum of the system overlaps in such way that it yields in third-order EP. Along this particular junction, the system is appeared to be protected by PT symmetry, where all of the three eigenvalues coexist with a real value and zero imaginary part. However, on both sides of the EP axis of eigenvalues, the system is in PT broken region, which means that the eigenvalues become complex. In Fig.\ref{fig2}, one can note at $ G_{a}=0 $, all eigenvalues have zero imaginary parts making them real. But it also means that the subsystem of the magnomechanics is not interacting with each other and system is disintegrated resulting in trivial Hermitian eigenfrequency state. It is interesting to note that here the value of non-Hermitian parameter is also equal to the magnon-photon coupling, i.e. $G_{a}=\Gamma=1$. If $G_{a}\ne\Gamma$, then the PT will be broken, because eigenvalues will have both real and imaginary parts. Further, the incident angle $\theta$, with the cavity axis, also plays a crucial part in making the system PT symmetric. But as discussed in reference \cite{Yasir_2024}, the system will only PT symmetric on $\pi/2$ and $\pi$. So, here in our calculation which are specifically for the understanding of steady-state dynamics, we choose $\theta=\pi$.

\section{steady-state Analysis}\label{std}
By using semi-quantum mechanical approach and treating operators of associated subsystem as classical variables, we put the time derivative equal to zero in Langevin equations and govern the steady-state solutions for the associate subsystem, reading as,  
\begin{eqnarray}
a_{s}&=&\dfrac{-(G_{a}-\mathrm{i}\Gamma \mathrm{e}^{\mathrm{i}\theta })m_{s}}{\Delta_{c}-\mathrm{i}\kappa_{a}},\label{5a}\\ 
m_{s}&=&\dfrac{-[G_{a}a_{s}-\mathrm{i}(\Gamma \mathrm{e}^{\mathrm{i}\theta } a_{s}-\eta) ]}{(\mathcal{A}-\mathrm{i}\kappa_{m} )},\label{5} 
\end{eqnarray}
where $\mathcal{A}=\Delta_{m}-2 G_{b}^{2}\omega_{b}/(\omega_{b}^{2}+\kappa_{b}^{2})\left | m_{s} \right |^2$ contains the magnonic detuning and the influences of the magnomechanical interaction \cite{Li_2018_Magnon}. Here it should be noted that, as previously mentioned, we will be focusing on the steady-state behavior of magnon and photons, so, we included the influences of phononic coupling and steady-states to above equation in order to simplify the calculations.
From above solutions, we get the steady-state cavity magnon numbers $\left | m_{s} \right |^2$ and photons numbers $\left | a_{s} \right |^2$, respectively, reading as,
\begin{eqnarray}
\left | m_{s} \right |^2&=& \frac{\eta^2(\Delta_{c}^2+\kappa_{a}^2)}{\mathcal{X}_m(\alpha)\mathcal{X}^*_m(\alpha) }, \label{magnon_number}\\
\left | a_{s} \right |^2&=& \frac{\eta^2(G_a^2+\Gamma^2\mathrm{e}^{\mathrm{i}2\theta })}{\mathcal{X}_a(\beta)\mathcal{X}^*_a(\beta) },\label{photon_number}
\end{eqnarray}
where,
\begin{eqnarray}
\mathcal{X}_m(\alpha)&=&\alpha-(G_{a}+i\Gamma e^{i\theta})^2, \\
\mathcal{X}_a(\beta)&=&\beta-(G_{a}+i\Gamma e^{i\theta})^2,
\end{eqnarray}
and,
\begin{eqnarray}
	\alpha&=&(\mathcal{A}-\mathrm{i}\kappa_{m})(\Delta_{c}-\mathrm{i}\kappa_{a}), \\
	\beta&=&(\mathcal{B}-\mathrm{i}\kappa_{m})(\Delta_{c}-\mathrm{i}\kappa_{a}),\\
	\mathcal{A}&=&\Delta_{m}-\frac{2 G_{b}^{2}\omega_{b}}{\omega_{b}^{2}+\kappa_{b}^{2}}\left | m_{s} \right |^2,\\
	\mathcal{B}&=&\Delta_{m}-\frac{2 G_{b}^{2}\omega_{b}(\Delta_{c}^2+\kappa_{a}^2)}{(\omega_{b}^{2}+\kappa_{b}^{2})(G_{a}^2+\Gamma^2 e^{i2\theta})}\left | a_{s} \right |^2.
\end{eqnarray}
Following steady-state analysis is done based on these calculations.

\subsection{Magnonic and Photonic Bi-stability}\label{sec4}
\begin{figure}[tp]
	\includegraphics[width=8cm]{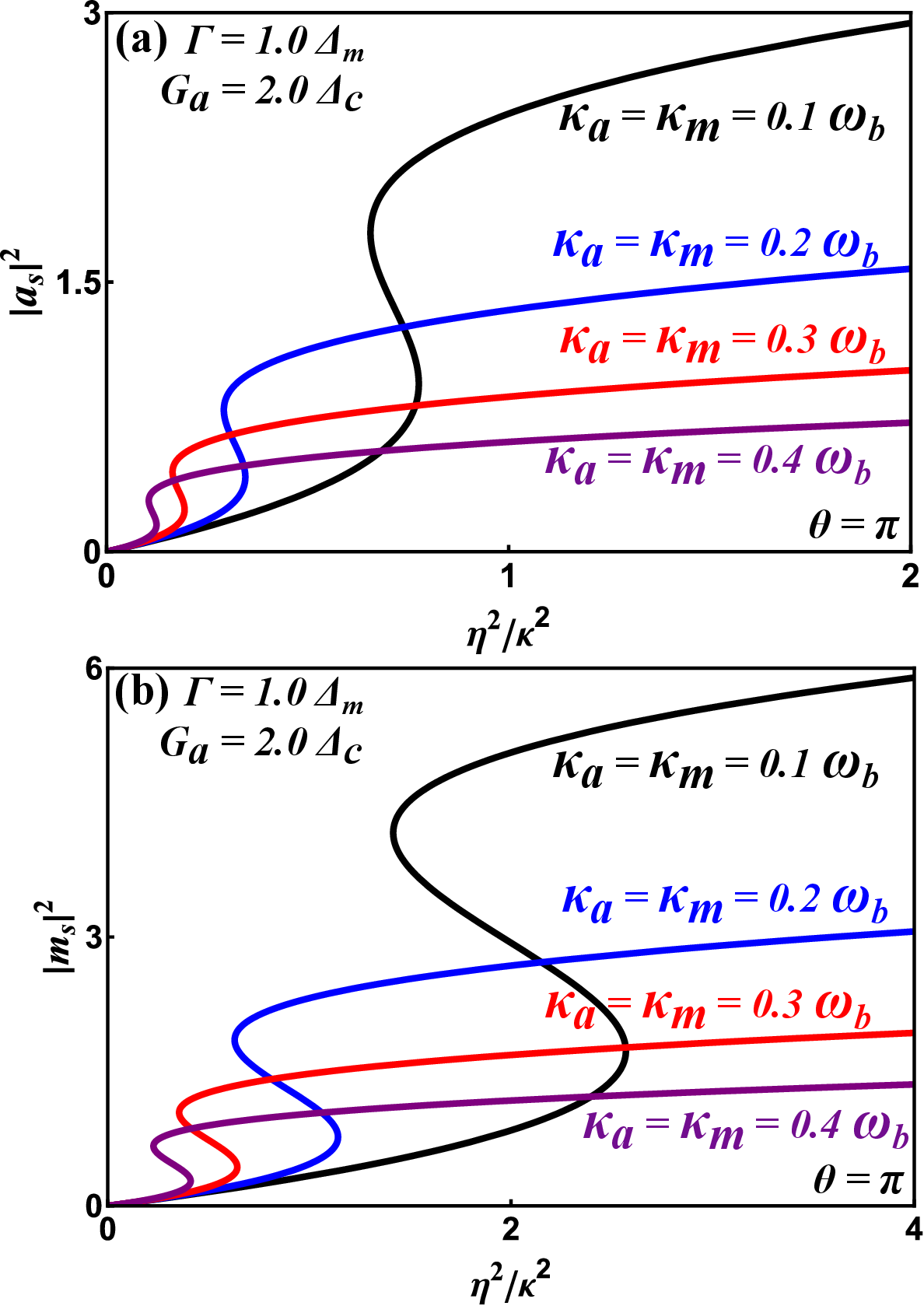}
	\caption{(a-b)The steady-state cavity photon number $\left | a_{s} \right |^2$ and magnon number $\left | m_{s} \right |^2$ as a function of the normalized driven magnetic field $\eta^2/\kappa^2$ for various values of dissipation of photon $k_{a}$ and magnon $k_{m}$ at constant $G_{a} = 1.5\Delta_{m}$ and  $\Gamma = 0.0\Delta_{m}$. While the other parameters are the same as used in Fig.\ref{fig2}.}
	\label{fig4}
\end{figure}

 \begin{figure*}[tp]
	\includegraphics[width=15cm]{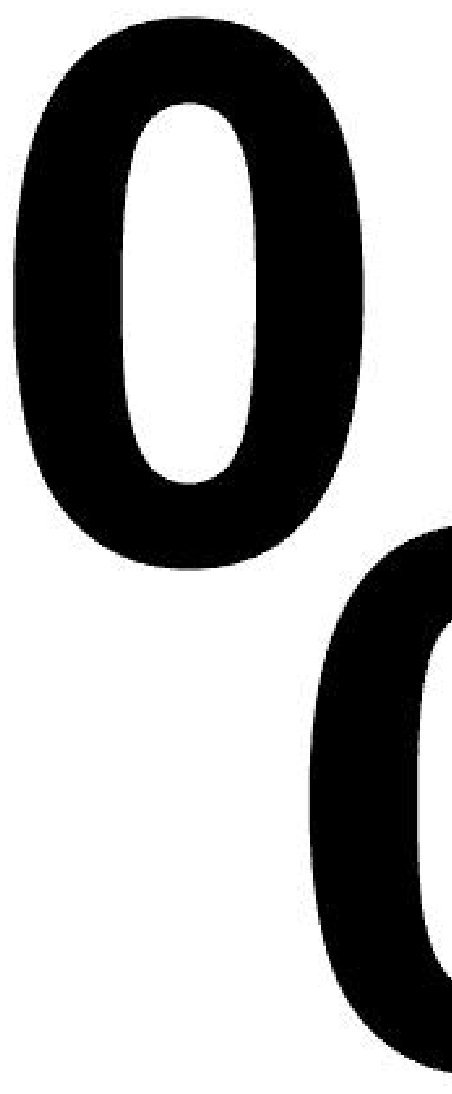}
	\caption{The contour representation of steady-state effective potential as a function of cavity photon numbers $\left | a_{s} \right |^2$ and magnon numbers $\left | m_{s} \right |^2$ with varying non-Hermitian parameter, $\Gamma = 0.0\Delta_{c}$ (a), $\Gamma = 0.5\Delta_{c}$ (b), $\Gamma = 2.0\Delta_{c}$ (c), $\Gamma = 1.5\Delta_{c}$ (d), $\Gamma = 2.0\Delta_{c}$ (e), and $\Gamma = 2.5\Delta_{c}$ (f) at constant $G_a= 2.0\Delta_{c}$. Other parameters are the same as in Fig. \ref{fig2}. } 
	\label{fig5}
\end{figure*}

The steady-state behavior of cavity photon numbers $\left | a_{s} \right |^2$ and magnon numbers $\left | m_{s} \right |^2$ is illustrated in Fig. \ref{fig3}, as a function of the normalized external magnetic field $\eta^2/\kappa^2$, at various values of coupling parameter $G_{a}$ (Figs. \ref{fig3}(a) and \ref{fig3}(b)) and non-Hermitian parameter $\Gamma$ (Figs. \ref{fig3}(c) and \ref{fig3}(d)). The occurrence of bistability (means with two stable states and one unstable state) can be seen versus $\eta^2/\kappa^2$. The photonic bistability increases with the increase in interaction between photons and magnons when  $G_{a}$ varies from  $G_{a} = 0.0\Delta_{c}$, $G_{a} = 0.8\Delta_{c}$, $G_{a} = 1.6 \Delta_{c}$, to $G_{a} = 2.4\Delta_{c}$ at constant non-Hermitian parameter $\Gamma =1.0\Delta_{m}$. However, at $G_{a} = 0.0\Delta_{c}$, and $G_{a}=0.8\Delta_{c}$, the photon numbers show no bistability and always possesses stable single stable state, which appears to be increasing with increase in $\eta^2/\kappa^2$. In other words, cavity photons are always stable with $\eta^2/\kappa^2$. 
  
But, when we extend the value of the magnon-photon coupling to $G_{a}= 1.6\Delta_{c}$  and  $G_{a}=2.4\Delta_{c}$, the photon numbers show bistable behavior. It is clear from the equation $(\ref{5a})$, that when the values of the coupling parameter are less than the non-Hermitian parameter, the system is in the gain condition i.e. $\frac{G_{a}}{\Gamma}<1$, but when we increase the value of the magnon-photon coupling gradually to $G_{a} = \Gamma$, i.e. $\frac{G_{a}}{\Gamma}= 1$, the system coalesces and we get no result. Such a condition is referred to EP \cite{Zhucheng_2020_PT,Wang_2023_Exceptional,Zhitao_2024}, as discussed in detail in the system description. All system eigenvalues coexist at a single point, known as singularity, will produce neither gain broken nor loss broken cavity photons forming a unconventional scenario with no excitation of cavity photons with external drive, which is indeed crucial and unlike previous opto/magnomechanical systems. Further increase in the magnon-photon from the non-Hermitian parameter (i.e. when $\frac{G_{a}}{\Gamma}> 1$) will result in the system shifting from a gain to a lossy broken condition, which will force the cavity photons to show bistability.

For steady-state behavior of magnon number, the magnon-photon coupling $G_{a}$ is appeared to be decreasing magnon number with increase in its values, as can be seen in Fig.\ref{fig3}(b). But, unlike the $G_{a}$ effects on $\left | a_{s} \right |^2$, $\left | m_{s} \right |^2$ illustrates bistability for all values of $G_{a}$, independent of whether it higher than $\Gamma$ or not. It is because the PT-symmetry behavior is occurring for cavity photon number and the previously mentioned gain and loss broken configurations are for cavity photons and not for steady-state magnons. At constant non-Hermitian parameter $\Gamma = 1.0\Delta_{m}$, the coupling parameter has the opposite effect on the bistable states phenomenon of the magnon numbers. The black solid line exhibits bistable behavior, but when we increase the coupling parameter from $G_{a} = 0.0\Delta_{c}$, $G_{a} = 0.8\Delta_{c}$, $G_{a} = 1.6\Delta_{c}$ to $G_{a} = 2.4\Delta_{c}$, the bistability gradually decreases. It is due to increasing photons and magnons interactions resulting in energy transfer from magnons to photons causing decrease in magnon numbers. In other words, photon scattering increases with increase in magnon-photon coupling, which consequently decreases magnon bistability.

Figure \ref{fig3}(c), shows the steady-state cavity photon numbers for the case when magnon-photon coupling is fixed to $G_a = 1.5\Delta_c$, and the non-Hermitian parameter is varying. The effect of the non-Hermitian parameter on photon numbers is opposite to that of the magnon-photon coupling parameter. Initially, when the value of the non-Hermitian parameter is less than the magnon-photon coupling (i.e $\Gamma =0.0\Delta_{m}$, $\Gamma = 0.6\Delta_{m}$, and $\Gamma = 1.2\Delta_{m}$) the steady-state photon number shows bistable nature. When we further increase the value of the non-Hermitian parameter i.e., $\Gamma = 2.4\Delta_{m}$, no stability. The increase in the non-Hermitian parameter results in gain behavior, which leads to cavity photon single stability state. As we know, bistability is result of system mediated nonlinearities, which get suppressed in gain configuration. Therefore, as a result, we will not bistability when $\Gamma$ is higher in magnon-photon coupling.

Similarly, Fig.\ref{fig3}(d) illustrates steady-state cavity magnon numbers against a normalized driven magnetic field at a constant coupling parameter $G_a = 1.5\Delta_c$ and varying non-Hermitian parameter. When the non-Hermitian parameter increases, the magnonic bistability similarly decreasing as with $G_{a}$ but towards higher values of $\eta^2/\kappa^2$. The reason is same as with $G_{a}$, $\Gamma$ also increases photon-magnon interaction in a way that cause increasing photon scattering to the cavity. Therefore, energy is transferred from magnons to photons growes, which in turn causes the number of magnons to be reduced resulting magnon bistability decreases.

\subsection{Dissipation Effects}
As in our case, hybrid magnomechanical system is an open quantum system, so, it is also important to see the effects of associated dissipation, i.e. $\kappa_{a}$ and $\kappa_{m}$ on the steady-state dynamics. Figs.\ref{fig4}(a) and \ref{fig4}(b) demonstrate such effects of photons ($k_{a}$) and magnons($k_{m}$) dissipation rate on the bistability of the steady-state cavity photon and magnon numbers at constant non-Hermitian parameter $\Gamma = 1.0\Delta_{m}$ and magnon-photon parameter $G_{a} = 2.0\Delta_{m}$. We observe that as the dissipation ratio increases gradually from $\frac{k_{a}}{k_{m}} = 0.1\omega_{b}$ to $\frac{k_{a}}{k_{m}} =0.4\omega_{b}$, both photons and magnon numbers become less bistable. Or, one can say, both photonic $\left | a_{s} \right |^2$ as well as magnonic $\left | m_{s} \right |^2$ are being suppressed with increase in associated system dissipation ratios. The increase in dissipation ratios lead to the rapid loss of photonic and magnonic energies, which consequently cause the suppression of photon and magnon number in all stable states resulting in overall decrease in bistability of photons and magnons. Thus, one can conclude that the steady-state cavity photon number as well as magnon number possess bistable state with respect to external magnetic field source, and that bistable behavior crucially depends on the gain and loss broken configuration of the system, unlike any previous investigation.

\subsection{Steady-State Effective Potential}
In order to further explore the steady-state dynamics, we plotted effective system steady-state potential versus the cavity photon number $\left | a_{s} \right |^2$ and magnon number $\left | m_{s} \right |^2$. Figs.\ref{fig5} (a-f) illustrate the relation between the photon numbers $\left | a_{s} \right |^2$ and magnon numbers $\left | m_{s} \right |^2$ against various non-Hermitian parameter, $\Gamma = 0\Delta_{m}$, $\Gamma = 0.5\Delta_{m}$, $\Gamma =1.0\Delta_m$, $\Gamma = 1.5\Delta_{m}$, $\Gamma=2.5\Delta_{m}$, and $\Gamma =3.0\Delta_m$, respectively, at constant coupling parameter $G_{a} = 2.0\Delta_c$. 
\begin{figure}[tp]
	\includegraphics[width=7cm]{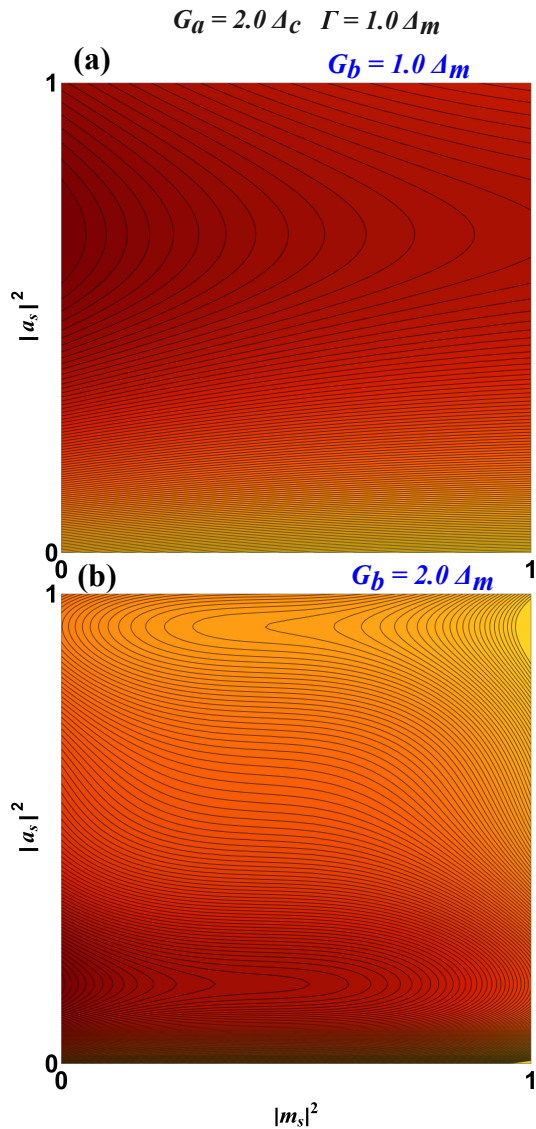}
	\caption{The contour representation of steady-state effective potential as a function of cavity photon numbers $\left | a_{s} \right |^2$ and magnon numbers $\left | m_{s} \right |^2$  with magnons-phonons coupling $G_b = 1.0\Delta_{m}$ (a) and $G_b = 2.0\Delta_{m}$ (b), at constant coupling parameter $G_{a}= 2.0\Delta_c$ and non-Hermitian parameter $\Gamma= 1.0\Delta_{m}$. While the other parameters are the same as used in Fig. \ref{fig2}. }
	\label{fig6}
\end{figure}

When the non-Hermitian parameter between magnons and photons is zero i.e. $\Gamma = 0\Delta_{m}$, the dissipative magno-optical cavity system clearly exhibits bistability due to maximal interaction between magnons and photons, which basically represents Hermitian scenario of the system. The bistability can be seen with nonlinear interactions between contour trajectories forming a combination of stable region and unstable points (sandal points). However, when the non-Hermitian strength increases gradually from $\Gamma =0\Delta_{m}$ to $\Gamma =0.5\Delta_{m}$, to $\Gamma =1.5\Delta_m$ energy exchange between photons and magnons increases, leading to a decrease in bistability. However the system is in a lossy broken state, so the system remains in bistability domain. It can be seen with nonlinear trajectory behavior of contours. However, at EP i.e. $\Gamma= 2.0\Delta_{m}$, the system will collapse and we will get nothing, similarly as previously discussed. Further increase in the non-Hermitian parameter, i.e., $\Gamma =2.5\Delta_{m}$,  and $\Gamma= 3.0\Delta_{m}$, yields in complete suppression of upper stable state or bistability and we will just have single state state, as can be seen in Figs.\ref{fig4}(e) and \ref{fig4}(f). During this transition, the system moves from a lossy broken state to a gain state. Thus, as the non-Hermitian parameter increases, energy transformation between magnons and photons occurs, resulting in increased energy of photons and, consequently, causing the system to transition from a bistable (Fig.\ref{fig5}(a)) to single stable state (Fig. \ref{fig5}(f)).

It is also important explore the effects of magnon-phonon coupling $G_b$ on the steady-state behavior of the system. Therefore, we also plot the steady-state effective system potential versus steady-state cavity photons and magnons number, keeping the values of the magnon-photon coupling $G_{a} = 2.0\Delta_{c}$, and the non-Hermitian parameter $\Gamma =1.0\Delta_{m}$ constants. It is observed that changing the values of the magnon-phonon coupling parameter from $G_{b}=1.0\Delta_{m}$ to $G_{b}=2.0\Delta_{m}$, results in a significant change in the bistability of the cavity photon and magnon numbers. At $G_{b}=1.0\Delta_{m}$ the steady-state potential shows stability (Fig. \ref{fig6}(a)), which can be seen with nonlinear interaction between contour trajectories. However, if we increase $G_{b}=2.0\Delta_{m}$, the bistability effects get enhance, as can be seen in Fig\ref{fig6}(b). The reason is simple that higher magnon-phonon couplings results in strong optomechanical interactions or more nonlinearities yielding in increase in bistability.

\section{Conclusion}\label{sec5}

In conclusion, we explore the steady-state dynamics of a non-Hermitian magnomechanical cavity. We show that both steady-state photons and magnons possess bistability not only just depending upon associate system interaction, like magnon-photon coupling and magnon-phonon coupling, but it also crucially depends on the strength of traveling field, which, in fact, makes system non-Hermitian. The cavity scattered photon are only bistable when the system in is lossy broken configuration, which means traveling field is weaker than the magnon-photon coupling. On other hand, it will just contain a single stable state growing with external drive. Further, it is found that the photon and magnon numbers are influenced oppositely by the magnon-photon coupling and in a similar pattern by the non-Hermitian parameter. As the magnon-photon coupling increases, the bistability of photons (magnons) increases (decreases), while the bistability of photons and magnons decreased as the non-Hermitian parameter value increased. Furthermore, we illustrate that the steady-state effective potential possesses similar bistable behavior as the photons and magnons are showing. We illustrate bistability here with nonlinear interaction between contour trajectories, similarly depending upon the ratio between traveling field strength and magnon-photon coupling. Our results expand general understanding of the interaction between photons and magnons in a non-Hermitian PT-symmetric systems, with potential implications for the subjects of quantum information, magnetic spintronics, and optical switching devices.

\begin{acknowledgments}
	K.A.Y. acknowledges the support of Research Fund for International Young Scientists by NSFC under grant No. KYZ04Y22050, Zhejiang Normal University research funding under grant No. ZC304021914 and Zhejiang province postdoctoral research project under grant number ZC304021952. 
\end{acknowledgments}

\appendix\section{}
By considering the system as a supermode consisting of three modes \cite{Zhucheng_2020_PT} and using Eq.(\ref{QLEs}), we can construct an effective Hamiltonian  \cite{Wang_2023_Exceptional} as follows
\begin{equation}
\hat{\mathcal{H}}=\begin{pmatrix} \hat{a}^{\dagger } & \hat{m}^{\dagger}  & \hat{b}^{\dagger}
\end{pmatrix}
\underbrace{\begin{pmatrix}
\Delta_{c}-\mathrm{i}\kappa_{a}  & G_{a}-\mathrm{i}\Gamma \mathrm{e}^{\mathrm{i}\theta  } & 0\\
G_{a}-\mathrm{i}\Gamma \mathrm{e}^{\mathrm{i}\theta  } & \widetilde{\Delta}_{m}-\mathrm{i}\kappa_{m} & \widetilde{G}_{b}\\
0 & \widetilde{G}_{b}^{\ast} & \omega_{b}
\end{pmatrix}}_{\hat{H}_{eff}}
\begin{pmatrix}
\hat{a} \\
\hat{m} \\
\hat{b}
\end{pmatrix},
\end{equation}
By solving the determinant of $H_{eff}$, we can get the eigenenergy of the three supermodes is $\lambda$, more explicitly, the cubic equation \cite{Jing_2017} as follows 
	\begin{equation}
	\lambda^{3}+r \lambda^{2}+s\lambda+t=0, \label{Eigenenergy_Equation}
\end{equation}
where the coefficients are 
\begin{gather}\begin{split}
		r &= -(\Delta_{a} -\mathrm{i}\kappa_{a}+\widetilde{\Delta}_{m} -\mathrm{i}\kappa_{m}+\omega_{b}-\mathrm{i}\kappa_{b}) \\
		s &= -G_{a}^{2}+2\mathrm{i}\Gamma \mathrm{e}^{\mathrm{i}\theta}G_{a}-\mathrm{i}\Gamma^{2}\mathrm{e}^{\mathrm{i}2\theta}-\widetilde{G}_{b}^{\ast}\widetilde{G}_{b} \\
		&+[(\Delta_{a} -\mathrm{i}(\kappa_{a}+\kappa_{b})][\widetilde{\Delta}_{m} -\mathrm{i}(\kappa_{m}+\kappa_{b})]+\kappa_{b}^{2}  \\
		&+(\Delta_{a} -\mathrm{i}\kappa_{a}+\widetilde{\Delta}_{m} -\mathrm{i}\kappa_{m})\omega_{b} \\
		t &= \mathrm{i}\kappa_{b}\left [ -G_{a}^{2}+2\mathrm{i}\Gamma \mathrm{e}^{\mathrm{i}\theta}G_{a}-\mathrm{i}\Gamma^{2}\mathrm{e}^{\mathrm{i}2\theta}+(\Delta_{a} -\mathrm{i}\kappa_{a})(\widetilde{\Delta}_{m} -\mathrm{i}\kappa_{m})\right ]  \\
		&+\widetilde{G}_{b}^{\ast}\widetilde{G}_{b}(\Delta_{a} -\mathrm{i}\kappa_{a}), 
	\end{split}
\end{gather}
\par 
By solving Eq.(\ref{Eigenenergy_Equation}), we can get three eigenvalues, where $\lambda_{1}$ is the real root and $\lambda_{2,3}$ are the complex roots as follows

\begin{widetext}
	\begin{gather}
		\begin{split}
			\lambda_{1} &=-\frac{r}{3}-\frac{2^{1/3}(-r^2+3s)}{3(-2r^3 +9rs-27t+3\sqrt{3}\sqrt{-r^2 s^2+4s^3 +4r^3t-18rst+27t^2}  )^{1/3}}\\
			&+\frac{(-2r^3 +9rs-27t+3\sqrt{3}\sqrt{-r^2 s^2+4s^3 +4r^3t-18rst+27t^2}  )^{1/3}}{3\times 2^{1/3}}   ,
		\end{split} \\
		\begin{split}
			\lambda_{2,3}&=-\frac{r}{3}+\frac{(1\pm\mathrm{i}\sqrt{3} )(-r^2+3s)}{3\times 2^{2/3}(-2r^3 +9rs-27t+3\sqrt{3}\sqrt{-r^2 s^2+4s^3 +4r^3t-18rst+27t^2}  )^{1/3}} \\
			&-\frac{(1\mp \mathrm{i}\sqrt{3} )(-2r^3 +9rs-27t+3\sqrt{3}\sqrt{-r^2 s^2+4s^3 +4r^3t-18rst+27t^2}  )^{1/3}}{6\times 2^{1/3}}   ,
		\end{split}
	\end{gather}
\end{widetext}

\newpage
\bibliography{ref}

\end{document}